\documentclass[conference]{IEEEtran}
\IEEEoverridecommandlockouts
\usepackage{cite}
\usepackage{amsmath,amssymb,amsfonts}
\usepackage{algorithmic}
\usepackage{graphicx}
\usepackage{textcomp}
\usepackage{xcolor}
\def\BibTeX{{\rm B\kern-.05em{\sc i\kern-.025em b}\kern-.08em
    T\kern-.1667em\lower.7ex\hbox{E}\kern-.125emX}}
\begin{document}

\title{Local power estimation of neuromodulations using point process modeling \\
\thanks{The authors acknowledge the partial support of the NSF grant 1631759.}
}

\author{Shailaja Akella$^{1}$\thanks{$^{1}$ Department of Electrical and Computer Engineering, University of Florida, Gainesville, FL, USA}, Ali Mohebi$^{6}$\thanks{$^{6}$  Department of Neurology, University of California, San Francisco, CA, USA}, Kierstin Riels$^{2}$,
Andreas Keil$^{2}$\thanks{$^{2}$ Department of Psychology and Center for the Study of Emotion $\&$ Attention, University of Florida, Gainesville, FL, USA},
Karim Oweiss$^{1,3,4,5}$\thanks{$^{3}$ Department of Biomedical Engineering, University of Florida, Gainesville, FL, USA}
\thanks{$^{4}$ Department of Neurology, University of Florida, Gainesville, FL, USA},
Jose C. Principe$^{1,3}$
\thanks{$^{5}$ Department of Neuroscience, McKnight Brain Institute, University of Florida, Gainesville, FL, USA}
}

\maketitle

\begin{abstract}
Extracellular electrical potentials (EEP) recorded from the brain are an active manifestation of all cellular processes that propagate within a volume of brain tissue. A standard approach for their quantification are power spectral analyses methods that reflect the global distribution of signal power over frequency. However, these methods incorporate analysis windows to achieve locality and therefore, are limited by the inherent trade - off between time and frequency resolutions. In this paper, we present a novel approach to estimate local power more precisely at a resolution as high as the sampling frequency. Our methods are well grounded on established neurophysiology of the bio-signals where we model EEPs as comprising of two components: neuromodulations and background activity. A local measure of power, we call Marked Point Process (MPP) spectrogram, is then derived as a power - weighted intensity function of the point process for neuromodulations. We demonstrate our results on two datasets: 1) local field potentials recorded from the prefrontal cortex of 3 rats performing a working memory task and 2) EEPs recorded via electroencephalography from the visual cortex of human subjects performing a conditioned stimulus task. A detailed analysis of the power - specific marked features of neuromodulations confirm high correlation between power spectral density and power in neuromodulations establishing the aptness of MPP spectrogram as a finer measure of power where it is able to track local variations in power while preserving the global structure of signal power distribution.     
\end{abstract}


\section{Introduction}
Extracellular electric potentials (EEP) recorded from the scalp by means of electroencephalography (EEG) and, invasively via local field potentials (LFP) and electrocorticography (ECoG),  reflect the average spatiotemporal interactions between neuronal sub populations and therefore, constitute a measure scale of neural constructs indicative of more overt types of behavioral phenomenon. The importance of these emergent population level field potentials has been reflected at both the research and clinical levels where EEPs have enabled interpretation of complex behavioral mechanisms such as Parkinson's disease \cite{b1}, sleep and memory consolidation \cite{b2}, and spatial navigation \cite{b3}, among few. 

Field potentials are highly non - stationary signals, exhibiting continuous fluctuations between unpredictable chaotic stages and predictable oscillatory stages \cite{b4}. Transient ordered patterns in the oscillatory stages, also known as neuromodulations, are a direct consequence of synchronized synaptic interactions in the neuronal assemblies at the level of single neurons \cite{b5}. These neuromodulations are immediately evident as waxing and waning oscillations in the raw as well as filtered EEP traces. The highly complex chaotic stages then correspond to spontaneous background activity which are known to be characterized by a $1/f$ power spectrum. Walter J. Freeman further confirmed this two - component hypothesis, when he experimentally showed that neuronal networks at rest produce featureless activity with their amplitudes conforming to a Gaussian distribution; deviations from Normality were then observed during active stages in the form of stable neuromodulations \cite{b6}. 

Given the consequence of field potentials in understanding brain dynamics and their usefulness to the fields of engineering and medicine, the problem of knowledge extraction from field potentials has been widely addressed in literature \cite{b7}. A solution was found in the `frequency content' of EEPs that appropriately characterizes the neuronal oscillations; ergo, making time - frequency analysis an integral part of brain studies. The task of spectral quantification of EEPs is notably demanding pertaining to the complex dynamics of non - stationary EEPs where it is required that an apt quantification of time variation also accounts for relevant frequency components.   

Spectral analysis techniques are one of the most heavily exploited methods for feature extraction from field potentials. Although these methods seek to identify neuronal oscillations, they determine the global distribution of signal power across the frequency spectrum. Time - frequency decomposition models such as short - time Fourier Transforms (STFT), wavelets, etc build on piece-wise stationarity of signals while applying Fourier Transform techniques on short windowed segments of the signals in order to construct the signal power spectrum. However, window based methods are restricted by the infamous time - frequency trade - off which lower bounds the product of time and frequency resolutions \cite{b8}, where an appropriate representation of time variation compromises relevant frequency variation and vice - versa. 

In this paper, we present a model based local power estimation measure, we call MPP spectrogram, to capture finer time variations of neuromodulatory power using precise marked features of the oscillatory events. The markers are obtained from methods elucidated in our previous work \cite{b9, b10, b11}. For the current study, our main objective was to relate the local power estimation achieved with our methods to the conventional power estimated by power spectrum density (PSD) methods. To achieve this, firstly, we demonstrate that power in neuromodulations, as obtained from the marked point process (MPP), are highly correlated with the PSD in the corresponding band. For this, we employ correntropy based measures for better quantification of the inter-dependency between the two measures. This is important because when applying our high resolution methodology for a given band, we do not estimate the PSD; while we still compare our methods with the vast literature that uses PSD quantification. Finally, we go on to show the ability of MPP spectrogram as a measure that goes beyond the pervasive power spectral measures where it offers not just the global power distribution but also enables access to time information at a resolution as high as the sampling frequency. We present our results as tested on two datasets in the $\gamma$ (80 - 150 Hz) and $\beta$ (15 - 30 Hz) frequency ranges, respectively: 1) LFPs recorded from 3 rats performing a two - alternative, instructed delay choice task across different days (sessions), 2) EEG recorded from 20 subjects across 6 channels in the visual cortex while performing a conditioned stimulus task. 

\section{Methods}

\subsection{Transient Model for Neuromodulatory Events}

Deriving motivation from the two - component hypothesis and building upon the concepts of conventional cortical rhythms, we define single channel bandpassed EEP traces,  $\tilde{y}(t)$, as a linear combination of background noise, $n_{0}(t)$ and neuromodulations, $y(t)$, as shown in (\ref{eq1}). Further, using shot - noise model representation \cite{b12}, the neuromodulations are re-constructed as convolutions between weighted, shifted Kronecker delta functions ($\boldsymbol{\delta}$(t)) and corresponding impulse response of temporal filters representing `typical' neuromodulations obtained from the data, and summarized as the dictionary atoms in $D = \{ d_{\omega} \}_{\omega = 1}^{K}$ according to (\ref{eq2}). Here, $a_{i}^{\omega} $ and $\tau_{i}^{\omega} $ are the projection coefficient and time of occurrence, respectively of the $i^{th}$ neuromodulatory event described by the dictionary atom, $d_{i,\omega}$. 

\begin{eqnarray}
\tilde{y}(t) = n_{0}(t) + y(t) = n_{0}(t)  + \displaystyle \sum_{i=1}^{N} y_i(t),  \label{eq1}\\
y_i(t) = \displaystyle \sum_{m = -\infty}^{\infty} a_{i}^{\omega} \boldsymbol{\delta}(t - \tau_{i}^{\omega} - m)d_{i,\omega}(m). \label{eq2} 
\end{eqnarray}

However, to realize such a representation, it is required that the neuromodulations are isolated from the background activity. We achieve this, in our first phase, we call `denoising' where a correntropy based similarity vector robustly separates the two components leveraging on the dissimilarities in their intrinsic distributions and the inherent predominance of background activity in EEP traces \cite{b11, b15}. A threshold, $\kappa$, defined as the minimum norm of the putative events is calculated which delineates the search space for neuromodulations in the subsequent phases. 

The final phase follows an unsupervised learning framework paralleling a K - means clustering algorithm to learn the dictionary, $D$, representative of `typical' neuromodulations from the EEP trace. Alternations between updation of the dictionary atoms and sparse decomposition, then, constitute the model training. The dictionary atoms are updated using single value decomposition where a correntropy based cost function is optimized with the purpose of repressing outliers to avoid bias in the estimation of principal components. The latter step of sparse decomposition follows a traditional matching pursuit technique where the search space is outlined by the threshold, $\kappa$, obtained from the denoising phase. 

\begin{figure}[!h]
\centering
\includegraphics[scale = 0.56]{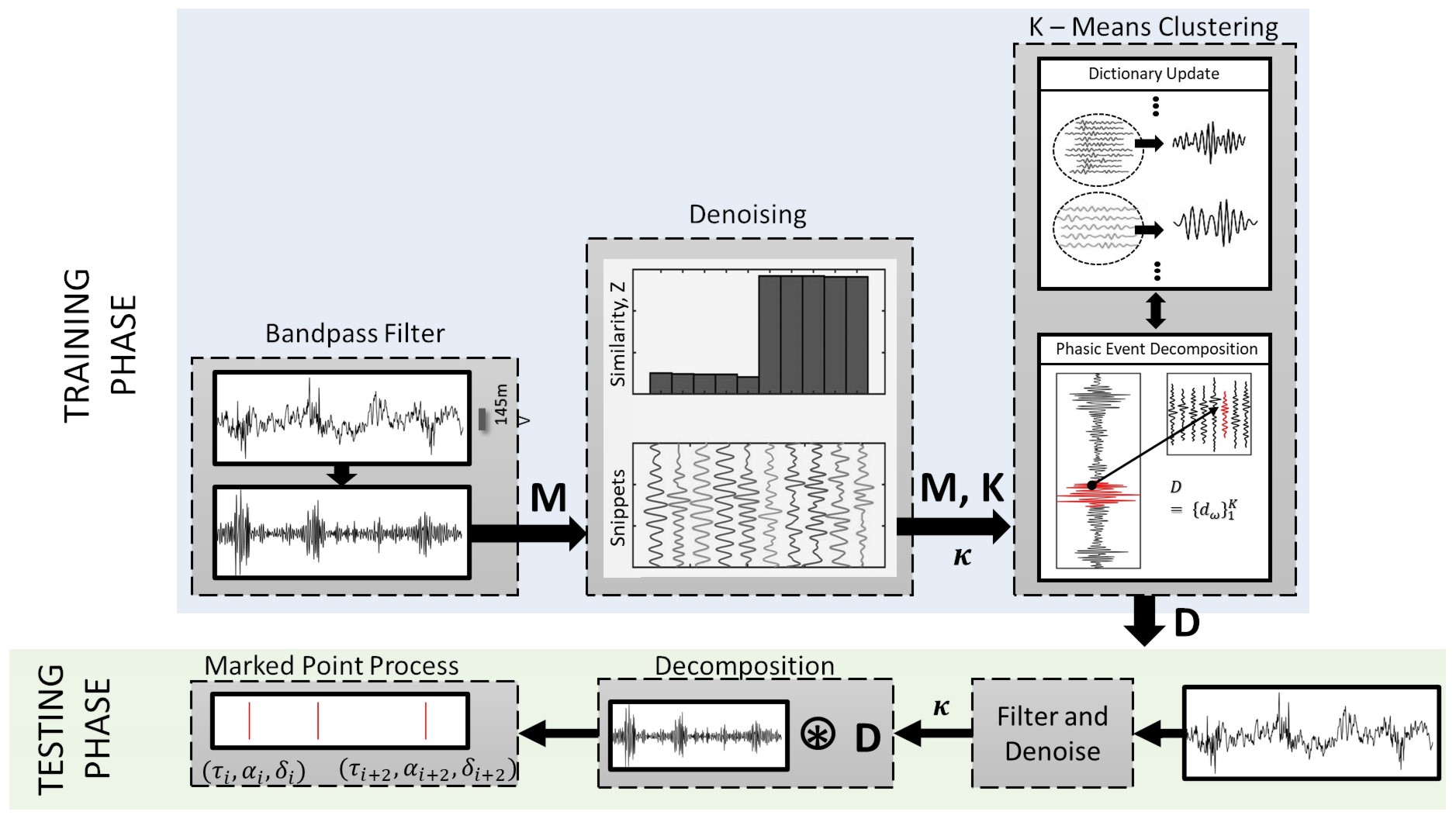}
\caption{Generative model. Training (top): After bandpass filtering the input EEP trace, the denoising phase exploits a correntropy - based similarity measure to calculate the threshold, $\kappa$. Following this, the learning framework estimates the bases vectors, D, via alternations between dictionary update and event decomposition phases. Testing (bottom): Input EEP trace is bandpassed and denoised to remove background noise. A final decomposition using the dictionary summarizes the features of each neuromodulation as a marked point process.}
\label{Model Structure}
\end{figure}

The model training is terminated when convergence is achieved in terms of an upper bound on the Frobenius difference norm between dictionaries estimated in successive iterations, or additionally, if a certain number of alternating optimization steps have been completed. The greedy approach of the methods demands that in order to avoid a local minima, the training be initialized with different initial seeds of the dictionary; wherefore, the final step in training includes determining the optimal dictionary with the maximum value of mutual coherence. Altogether, the model is able to learn representations of neuromodulations by adapting a data - driven approach while only depending on two hyperparameters: 1) maximum duration of neuromodulation, $M$ and 2) maximum number of dictionary atoms, $K$. A flowchart of the involved methods in the training phase is included in Fig. \ref{Model Structure} and for a more elaborate explanation of the model, we direct the reader to \cite{b9}.

\subsection{Neuromodulatory Features}
To this end, we have designed a dictionary prototypical of neuromodulations from a single channel EEP trace. Model testing then involves a simple convolution with the signal to obtain the markers of each neuromodulation, in terms of its time of occurence ($\tau_{k}$), maximum amplitude ($\alpha_{k}$) and duration ($\delta_{k}$) constructing the MPP, via the intensity function of the timings and the joint probability distribution of the neuromodulatory features, duration and amplitude (Fig.\ref{Model Structure}, bottom panel). Given complete access to each neuromodulatory event at time points, $\tau_{k}: k = 1,2, ... N$, we extend our feature space by defining local power in each neuromodulation, $P_\delta (\tau_{k})$ according to (\ref{eq3}).  A major advantage of $P_\delta (\tau_{k})$ is its ability to clearly isolate neuromodulatory power from power due to background noise. This property is emphasized in Fig. \ref{Spectrogram} where power spectrum obscures the boundary between background activity and neuromodulations due to their dependence on fixed windows, while $P_\delta (\tau_{k})$ is able to maintain distinct boundaries between the two EEP components.   

\begin{eqnarray}
P_\delta(\tau_k) =\frac{1}{\delta_{k}} \displaystyle \sum_{n = \frac{-\delta_k}{2}}^{\frac{\delta_k}{2}-1} \tilde{y}^{2}(\tau_{k} + n).  \label{eq3}
\end{eqnarray}

\begin{eqnarray}
\lambda_\alpha(t) = \displaystyle \sum_{n = -\infty}^{\infty}     \left[ \displaystyle \sum_{k = 1}^{N}  P_{\delta}(\tau_k) \boldsymbol{\delta}(n - \tau_k) \right] \kappa_\sigma(t - n)  \label{eq4}
\end{eqnarray}

\begin{eqnarray}
\rho_\delta = \frac{\displaystyle \sum_{k = 1}^{N}\delta_{k}}{L}   \label{eq5}
\end{eqnarray}

Further, we define MPP spectrogram ($\lambda_\alpha$) as the power weighted intensity function to capture the local variations in power due to $N$ detected neuromodulations from the EEP trace according to (\ref{eq4})  where $\kappa_\sigma$ is a Guassian kernel of bandwidth $\sigma$ and $\boldsymbol{\delta}(t)$ is the Kronecker delta function. It is important to note that, unlike power spectrum, $\lambda_\alpha$ is able to retain complete time information in the frequency range of interest while consistently limiting leakage from power due to background noise as it builds only on $P_\delta (\tau_{k})$.  

In addition to power measures, we also define a density measure, we call phasic event density ($\rho_\delta$), as the relative proportion of samples in the EEP trace of length $L$ that corresponds to neuromodulations. Phasic event density can be thought of as a measure that draws a parallel with the $l_0$ pseudonorm applied to the concept of sparse decomposition.  

\subsection{Correntropy Coefficient}

Throughout the paper, we inspect inter-dependencies between random variables via a correntropy measure, termed as correntropy coefficient, $\eta$ \cite{b15}. Specifically designed as a generalized correlation coefficient, the estimate builds on cross - correntropy as shown in (\ref{eq7}) for non - degenerate random variables $X  =\{ {x_i}\}_1^N$  and $Y = \{{y_i}\}_1^N$ where $\kappa_\sigma$ is a Gaussian kernel of bandwidth $\sigma$. 
The ``centered" cross - correntropy (\ref{eq6}) estimate is analogous to the conventional cross - covariance measure of the correlation coefficient where it is defined as the difference between the joint and the product of marginal expectations of $\kappa(X,Y)$; the centering is pertinent to an explicit extraction of the mean value. Normalizing the centered cross correntropy with respect to their (centered) autocorrentropies, then, completely defines the correntropy coefficient (\ref{eq7}).

\begin{eqnarray}
\hat{u}_\sigma(X,Y) = \frac{1}{N}\displaystyle \sum_{i = 1}^{N} \kappa_\sigma(x_i - y_i)  - \frac{1}{N^2}\displaystyle \sum_{i = 1} ^{N} \sum_{j = 1}^{N} \kappa_\sigma(x_i - y_j)  \label{eq6}
\end{eqnarray}

\begin{eqnarray}
\eta = \frac{\hat{u}_\sigma(X,Y)}{\sqrt{\hat{u}_\sigma(X,X) \hat{u}_\sigma(Y,Y)}}  \label{eq7}
\end{eqnarray}

The rationale for implementing a correntropy - based measure, as opposed to correlation, lies in its ability to reject outliers by controlling the kernel bandwidth, $\sigma$ \cite{b15}. In this scenario, outliers are associated with local artifacts in the data or the poor estimation of events by the MPP algorithm, that are although rare, distort the correlation fits too much. Finally, relation between the random variables under study was also further scrutinized via best fit lines obtained from a maximum correntropy criterion (MCC) based regression to conclude the analysis of any inter-dependency \cite{b15}. 

\section{Experimental Setting}
We tested our power measures on both LFP and EEG data to validate our two premises: 1) neuromodulations, as detected by the model, maximally contribute to signal power and 2) MPP spectrogram is a finer measure of power spectrum which also reflects the global distribution of signal power. This section presents the details of the two datasets analyzed.   

\subsection{Dataset-1}

Local field potentials were recorded from the dorsal prelimbic cortex (dPrL) across 3 rats performing a two-alternative, instructed delay forced choice task as per studies reported in \cite{b13}.  A microwire array was implanted in layer V of the prefrontal cortex in the right hemisphere such that electrodes 16 electrodes rested in the dPrL. The entire data was downsampled to 500 Hz for analysis. Further, all procedures involving animals were reviewed and approved by the Institutional Animal Care and Use Committee (IACUC).

The experiment was setup in an acoustically isolated operant conditioning chamber comprising of three nose poke holes: a fixation hole in the center was flanked by target holes on either side. Trials were self - initiated by the subject by positioning it's snout in the fixation hole. Before being presented with an instruction cue, the subject was required to place its nose in the fixation hole for 1s. The instruction cue was a variable pitch single tone delivered at 60 dB whose pitch determined the target hole; a low pitch tone corresponded to the right target hole, while a high pitch tone cued the left target hole. The instruction cue initiated a delay period, the length of which was chosen pseudo - randomly from a uniform distribution, $\textbf{U}(1.5, 2)$. A Go cue, consisting of a white noise auditory stimulus delivered at 60 dB, then indicated the end of the delay period following which the subject was required to place its nose in the appropriate target hole. Visits to the instructed targets were rewarded immediately by delivering a 45mg food pellet (Bio-Serv, NJ), while incorrect visits were unrewarded. Fig.\ref{Trial Structure}\textbf{A} presents the trial structure for the experiment. A total of 8 sessions, that is, 1 session by subject 1, 3 sessions by subject 2 and 4 sessions by subject 3 were analyzed.

\subsection{Dataset-2}

\begin{figure}[!ht]
\centering
\includegraphics[scale = 0.12]{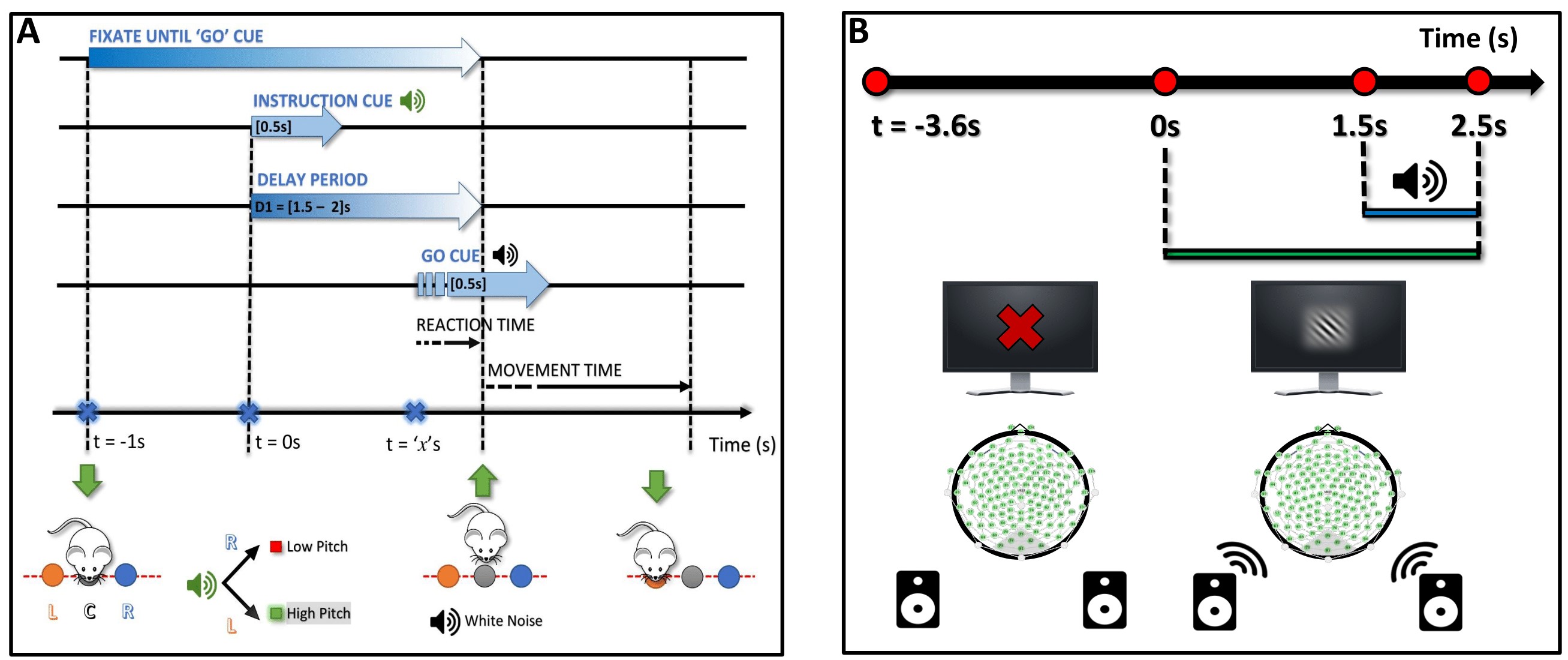}
\caption{Time course of the tasks. \textbf{A} LFPs recorded from rat prefrontal cortex. Gradients arrows represent variable time periods. The cues last for 0.5s each. \textbf{B} EEG data recorded from the human visual cortex. The epochs analyzed included the period [-3.6, 1.5]s. }
\label{Trial Structure}
\end{figure}

EEG data was recorded from 20 subjects while performing a task where an unconditioned stimulus was randomly interleaved with a conditioned stimulus in each trial. A total of 129 channels were recorded, of which 6 channels in the visual cortex were chosen for the current analysis (shaded channels, Fig. \ref{Trial Structure}\textbf{B}). Before trial initiation, the subject was required to look at a fixation cross that was presented at the center of the screen and lasted $\sim3.6$ s. Each trial was initiated by presenting the conditioned stimuli (CS), a Gabor patch with a 1.5 degree left tilt and a Michelson contrast of 0.63, at the center of the screen. The CS was displayed for a total of 2.5s throughout the trial. The unconditioned stimulus (UCS), randomly paired with the CS in $\sim50\%$ of the trials, was a 96 dB sound delivered via 2 computer speakers positioned behind the participant. The UCS lasted for 1s and was set about 1.5s after trial initiation.

Epochs analyzed in the current study were 5.1s in duration which included a 3.6s long display of a fixation cross in the center of the screen prior to trial initiation and 1.5s after CS onset as shown in Fig. \ref{Trial Structure}\textbf{B}. All data was sampled at 500 Hz and averaged to a single reference. The study was approved by the local Institutional Review Board. 

\begin{figure}[!ht]
\centering
\includegraphics[scale = 0.23]{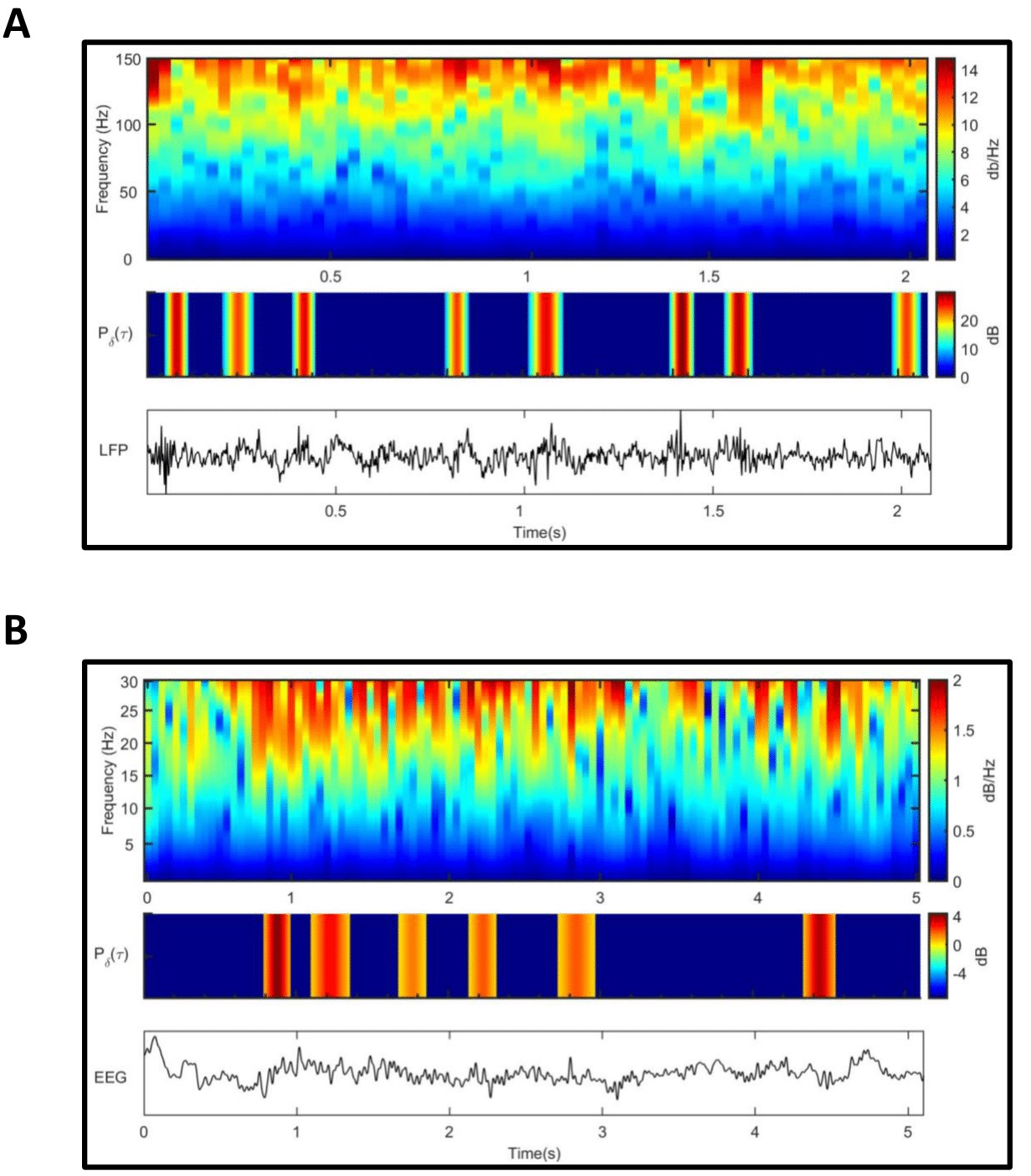}
\caption{Power represented via $P_\delta$ compared to the power spectrum. In each sub-figure, top panel presents the spectrogram of the recorded field potenials, middle panel corresponds to power in neuromodulations evaluated via $P_\delta$ in the frequency band under study and the bottom panel plots the raw field potentials throughout the trial. \textbf{A} Dataset 1: $\gamma$ (80 - 150 Hz); Session 1, channel 2, Trial 12. \textbf{B} Dataset 2: $\beta$ (15 - 30 Hz); Subject 1, channel 2, Trial 24.}
\label{Spectrogram}
\end{figure}

\section{Results}
Using the generative model for neuromodulatory events, we learnt the markers of $\gamma$ neuromodulations from the LFPs of the rat subjects and of $\beta$ neuromodulations from EEG data recorded from human subjects. The bandpassed traces in each frequency band were obtained using FIR filters of quality factor, Q $\sim1.5$ (center frequencies - $\gamma$: 115 Hz, $\beta$: 22.5 Hz; order - $\gamma$: 43; $\beta$: 54). After thorough visual analysis of the EEP traces and taking into account the number of cycles of oscillations, maximum duration of $\gamma$ and $\beta$ neuromodulations, $M$, was set at 100 ms ($\sim$10 cycles)  and 300 ms ($\sim$7 cycles), respectively. Further, following similar analysis as in \cite{b9}, we upper bounded the number of dictionary atoms, $K$, to $30$ for LFPs and $60$ for the EEG data. Correntropy measures implemented throughout the model used the kernel bandwidths as determined by the Silverman's rule \cite{b14} for an appropriate projection into the joint space of the random variables involved. MPP spectrogram measure was obtained from the MPP representation using Gaussian kernels of bandwidth 80 and 100 ms for $\gamma$ and $\beta$ frequency bands, respectively. Finally, PSD measures throughout the analysis were calculated using STFT-Welch method with Gaussian window lengths set equal to 0.4 and 0.7s for LFPs ($\gamma$ frequencies) and EEG data ($\beta$ frequencies), respectively with 50$\%$ overlap between windows.  

We began our analysis via simply juxtaposing spectrograms of the EEP traces (STFT, overlap - 70 $\%$,  window lengths - $\gamma$: 0.2 s, $\beta$: 0.3 s) with their corresponding detections of neuromodulations represented as their power, $P_\delta(\tau_{k})$. Exemplary plots are shown in Fig.\ref{Spectrogram} where it is evident that neuromodulations are detected by the MPP model in that portion of the signal where the PSD estimate (in the specified frequency band) is large; importantly, our method is able to precisely pin point the time of the event that is contributing to this increase in the PSD. Such an apt representation of neuromodulations by $P_\delta(\tau_{k})$  further motivated our analysis. Again, it is important to note the distinct boundaries evident between the neuromodulations and background activity as captured by $P_\delta(\tau_{k})$, which are absent in the spectrogram of the signal.  

In order for neuromodulations to be representative of signal power, it would be required that the two measures share a positive correlation. This premise was tested for recordings from both data sets by evaluating the correntropy coefficients between the normalized PSD (nPSD), as in (\ref{eq8}) and total power in neuromodulations, $\Sigma P_\delta$,  in the specified frequency band for each trial. Such normalization of the PSD was required in order to account for the difference in the units of the two measures: that is, while PSD is calculated over a window of fixed length,  $P_\delta$ calculates power only over the neuromodulatory event. Consequently, nPSD was normalized to represent power as captured by spectral analysis in the same duration as $P_\delta$ with the help of phasic event density, $\rho_\delta$, as in (\ref{eq8}) where $N_w$ is the number of windows used to calculate PSD.  

\begin{eqnarray}
nPSD = PSD*N_{w}*\rho_\delta. \label{eq8}
\end{eqnarray}

\begin{figure}[!ht]
\centering
\includegraphics[scale = 0.125]{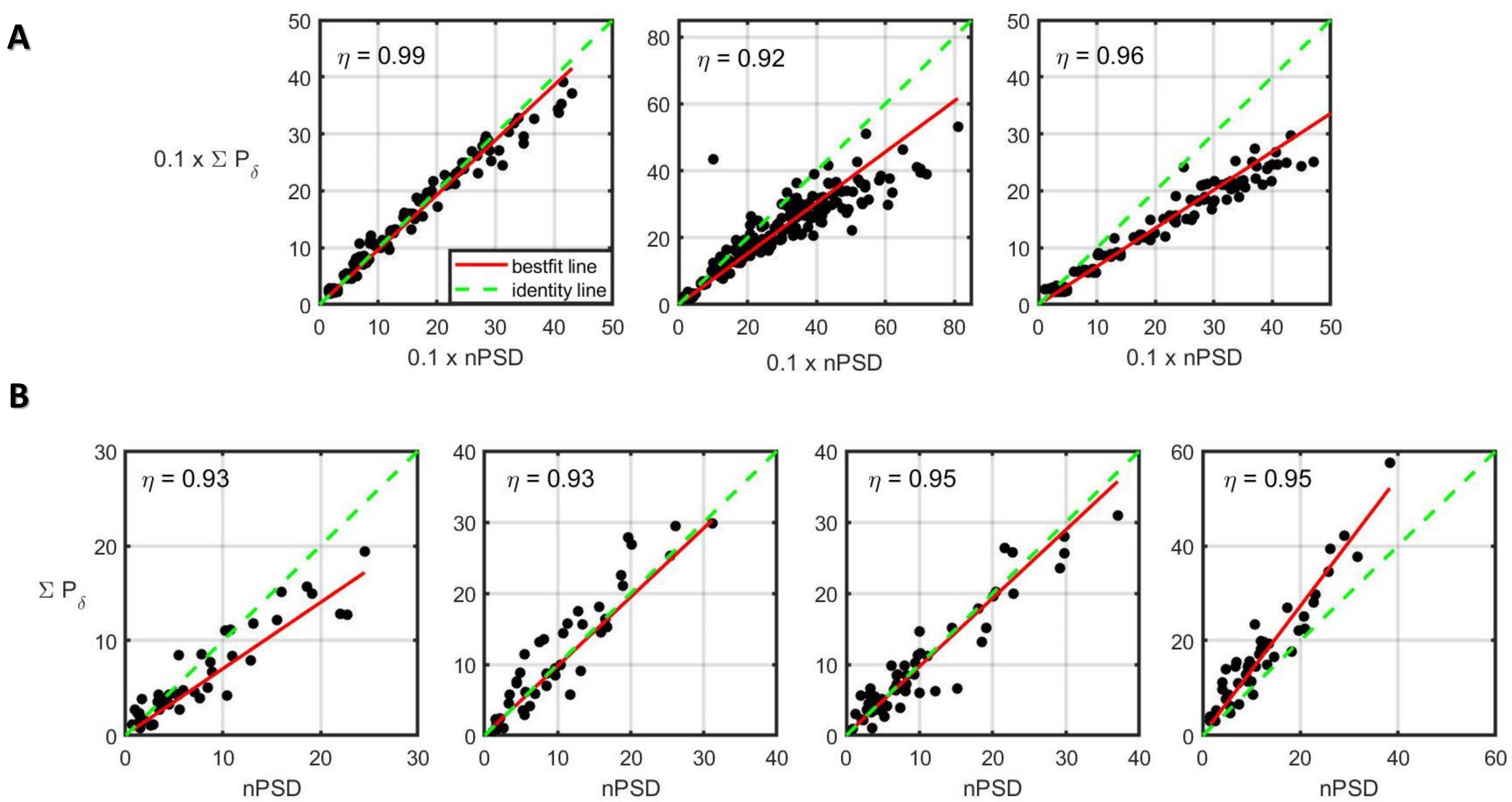}
\caption{Total power in neuromodulations plotted against nPSD in the corresponding frequency range for each trial. Each black marker corresponds to a trial and $\eta$ is the correntropy coefficient. \textbf{A} Dataset 1: $\gamma$ (80 - 150 Hz); (L - R) Subjects 1 - 3; Sessions - 1, 3, 2, respectively;  Channel 7,  \textbf{B} Dataset 2: $\beta$ (15 - 30 Hz); (L - R) Subjects 1, 4, 7 and 17; Channel 73.}
\label{PINvsnPSD}
\end{figure}

Results from across 3 sessions of LFP recordings corresponding to each rat and across 4 subjects' EEG data are presented in Fig.\ref{PINvsnPSD}\textbf{A} and \textbf{B}, respectively for a single channel. Moreover, best fit lines determined using   MCC further emphasize the interrelation between the two measures. It is worth noting that, in the plots, most of the detected neuromodulations exist along a line in the 2D space, but there are a few detections that are outliers. These outliers will bias the regression line and provide bogus correlation values. By contrast, correntopy is capable of aligning the regression line with the inherent trend, corresponding to a better estimation of the correlation between the two methods than linear regression does. Further, although the slope of the best fit line is very close to 1, in most cases, the slope was found to be slightly lower than that of the identity line. We attribute this to two plausible scenarios: 1) higher nPSD corresponding to power in background activity or 2) missed detections of neuromodulatory events. Scenario-1 reveals scope for analyses of brain processes that contribute to power in background activity; such analyses would merit future work.  

\begin{figure}[!ht]
\centering
\includegraphics[scale = 0.125]{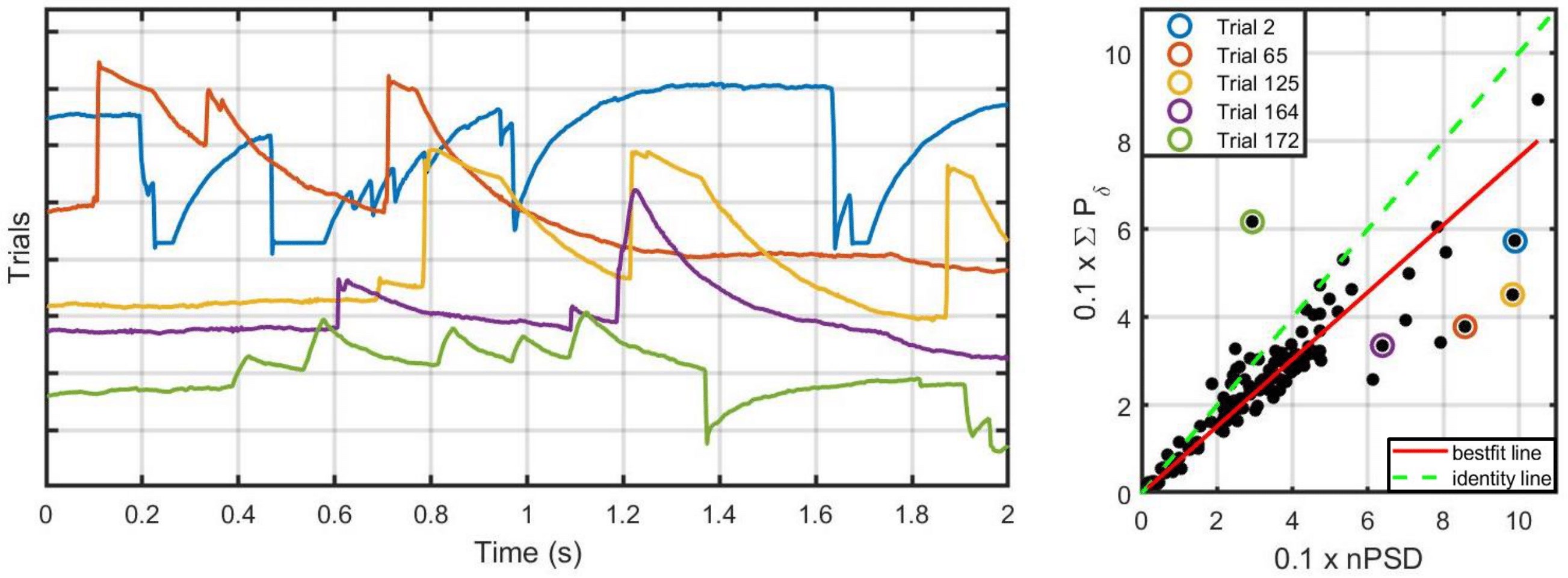}
\caption{Noisy signal detection. In the plot on the \textit{right}, $\Sigma P_\delta$ plotted against nPSD obtained from $\gamma$ neuromodulations for all trials in LFP recordings of a single channel from dataset-1 show large deviations from best fit line. Five outliers are mapped to their corresponding trial recordings on the \textit{left} plot depict artifact noise. Subject 2; Session 2; Channel 2.}
\label{badCorr}
\end{figure}

Additionally, these scatter plots not just serve as a validation for our methods, but also as a device to detect noisy recordings; for any large deviations from the best fit line produced by an appropriately tuned generative model would most likely correspond to artifact noise or bad channels in the recordings. This was observed in an LFP recording from dataset-1 where noise in the signal contributed to a more dispersed scatter plot (Fig. \ref{badCorr}). 

\begin{figure}[!ht]
\centering
\includegraphics[scale = 0.11]{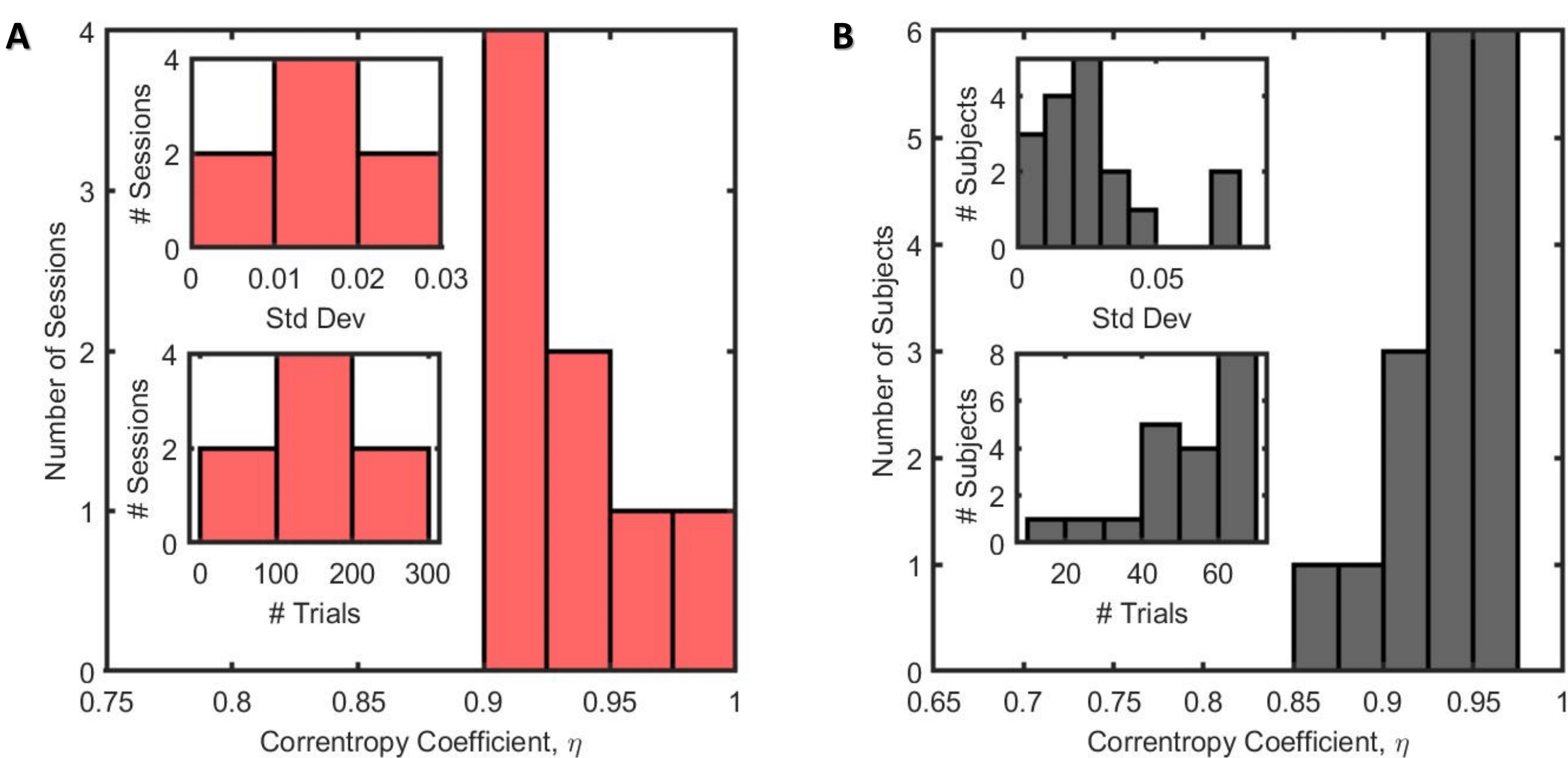}
\caption{Summary of analysis on inter-dependency between neuromodulations and signal power. Primary histograms summarize the mean correntropy coefficients between $\Sigma P_\delta$ and nPSD calculated across all channels for each session/subject analyzed. (Top inset) Histogram of standard deviations of coefficients. (Bottom inset) Histogram of number of trials performed in each session/by each subject. \textbf{A} Dataset 1: $\gamma$ (80 - 150 Hz); \textbf{B} Dataset 2: $\beta$ (15 - 30 Hz). }
\label{All_corrs}
\end{figure}

Mean correntropy coefficient ($\eta$) values across all channels for each LFP session and for each subject's EEG are summarized in Fig. \ref{All_corrs}, respectively, along with their standard deviations and trial details. High dependencies between power in neuromodulations and PSD confirms a high correlation between the measures. Moreover, normalization of PSD with phasic event density shows that neuromodulations maximally contribute to signal power. These results ratify our argument by bringing out a two - fold conclusion: higher number of neuromodulations not just imply a higher signal power but also that the signal power is most dominant in these neuromodulations. Finally, for the analysis of $\eta$, only 17 out of the total 20 subjects from dataset-2 were considered, as 3 subjects performed less than 30 trials and therefore, their results would not be representative of the inter - dependency between the variables under study.    

\begin{figure}[!ht]
\centering
\includegraphics[scale = 0.21]{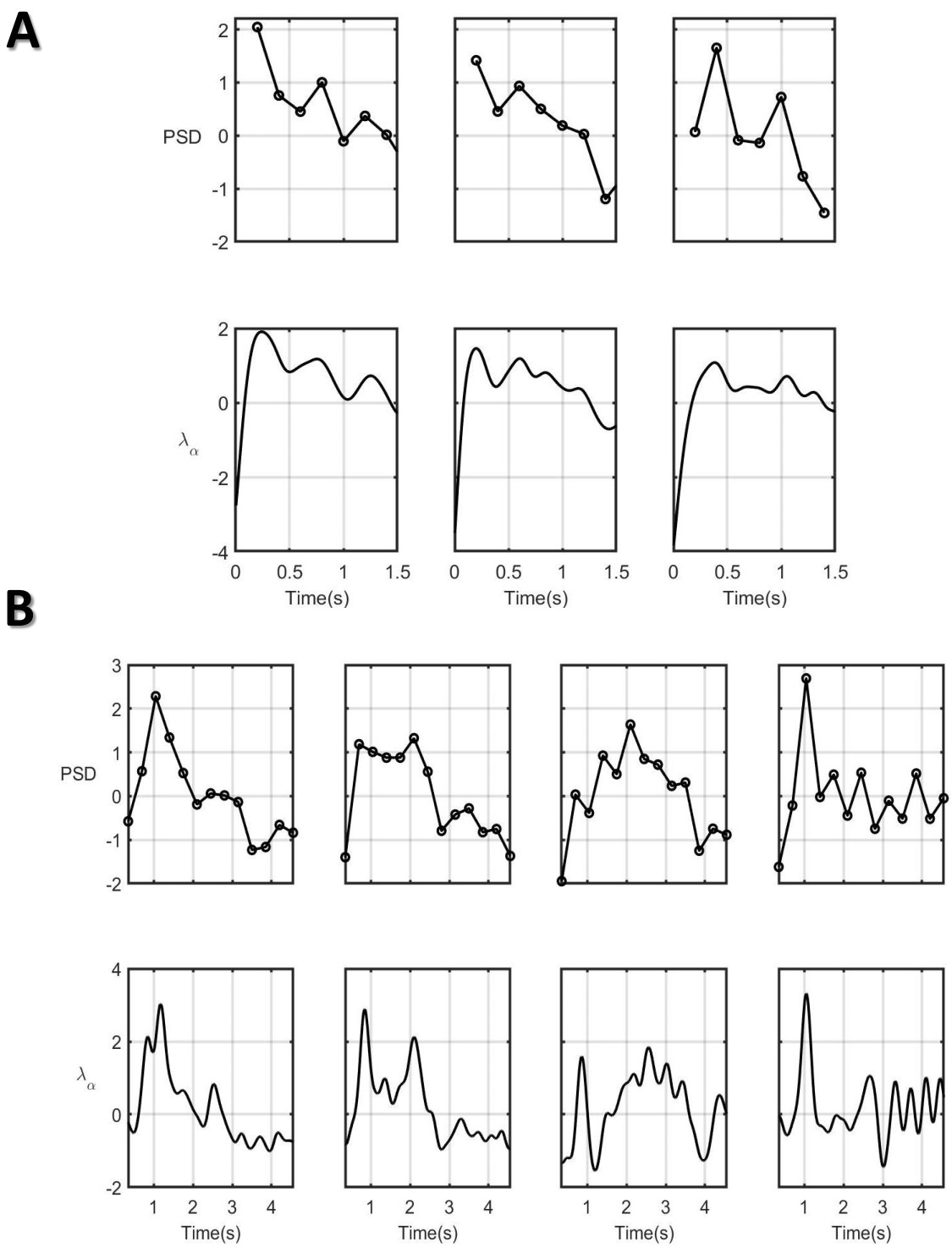}
\caption{Comparing power variations over time as captured by PSD and MPP spectrogram. In each subfigure, top panel correspond to the trial - averaged PSD and the bottom panels are trial - averaged $\lambda_\alpha$. Values of PSD and $\lambda_\alpha$ were normalized for visual purposes. \textbf{A} Dataset 1: $\gamma$ (80 - 150 Hz); (L - R) Subjects 1 - 3; Sessions - 1, 3, 2, respectively; \textbf{B} Dataset 2: $\beta$ (15 - 30 Hz); (L - R) Subjects 1, 4, 7 and 17}
\label{AM power}
\end{figure}

Having validated that neuromodulation power are a proxy for signal power as estimated via PSD in that frequency band, we were interested in assessing the ability of MPP spectrogram, which builds on power in these neuromodulations, to capture local variations in signal power. Such a representation would allow for greater access to time information at resolutions unachievable by other power spectral measures. For this, we evaluated the power spectrum of the signal using STFT and averaged the obtained spectrogram across the frequency band under study. These plots were then compared with those obtained using MPP spectrogram. Trial - averaged power spectrum plots juxtaposed with trial - averaged MPP spectrogram plots are presented in Fig. \ref{AM power}\textbf{A, B} corresponding to $\gamma$ and $\beta$ neuromodulations, respectively. Two observations are immediate from the figures that highlight the advantages of MPP spectrogram: 1) the general similarity in the variations of power over the trial period between the power spectrum and MPP spectrogram reflects the capability of MPP spectrogram to maintain the global power distribution of the signal, and 2) the finer details of variations that are only captured in the plots of MPP spectrogram. Therefore, MPP spectrogram is a high time - resolution power measure founded on concepts of neuromodulations that defines a finer, and more accurate representation of local as well as global signal power. Finally, it is worth noting that our methods do not lose any frequency resolution and are in fact a cumulative representation of \textit{all} frequencies in the specified bandwidth, i.e., there is no `sampling' of frequencies, unlike PSD methods. 

\section{Conclusion}

Through this paper, we introduce MPP spectrogram as a high - time resolution power measure obtained as a by - product of the neuromodulatory event detection model. The model's advantage lies in its ability to represent an EEP trace as a marked point process (MPP) by characterizing each neuromodulation in terms of its features - amplitude, duration and time of occurrence. We exploit this explicit access to neuromodulatory properties to expand the feature space to include power in neuromodulation ($P_\delta(\tau_k)$) as a clear - cut marker that distinguishes the power in neuromodulations from background power. Leveraging on the properties of $P_\delta(\tau_k)$, we construct MPP spectrogram to capture local variations in neuromodulatory power over time. Further, in order to validate the aptness of MPP spectrogram, we demonstrate firstly the dominance of neuromodulatory power in signal power using correntropy based techniques. We test our hypotheses on LFPs recorded from a rat prefrontal cortex and human EEG data recorded from the visual cortex. We show explicit variations in neuromodulatory power captured via MPP spectrogram as opposed to PSD, which, although closely resembles the MPP feature modulations, lacks the high time - resolution afforded by it.

\end{document}